\documentclass[aps,prb,manuscript]{revtex4}
\usepackage{graphicx}
\usepackage{epsfig}
\usepackage{color}
\usepackage{multirow}
\usepackage{dcolumn}
\usepackage{bm}
\usepackage{float}

\begin{document}
\draft
\title {Si-doped Defect in Monolayer Graphene: Magnetic Quantization}

\author{P. H. Shih$^{1}$, T. N. Do$^{2}$, B. L. Huang$^{1}$, G. Gumbs$^{3,4}$, D. Huang$^{5}$ and M. F. Lin$^{1,6}$}
\affiliation{$^{1}$Department of Physics, National Cheng Kung University, Tainan, Taiwan 701 \\
$^{2}$Institute of Physics, Academia Sinica, Taipei, Taiwan 115 \\
$^{3}$Department of Physics and Astronomy, Hunter College of the City University of New York, 
695 Park Avenue, New York, New York 10065, USA \\
$^{4}$Donostia International Physics Center (DIPC), P de Manuel Lardizabal, 4, 20018 San Sebastian, Basque Country, Spain \\
$^{5}$US Air Force Research Laboratory, Space Vehicles Directorate, Kirtland Air Force Base, New Mexico 87117, USA \\
$^{6}$Quantum Topology Center, National Cheng Kung University, Tainan, Taiwan 701
}

\date{\today}

\begin{abstract}
We explore the rich and unique magnetic quantization of Si-doped graphene defect systems with various concentrations and configurations using the generalized tight-binding model. This model takes into account simultaneously the non-uniform bond lengths, site energies and hopping integrals, and a uniform perpendicular magnetic field (${B_z\hat z}$).
The magnetic quantized Landau levels (LLs) are classified into four different kinds based on the probability distributions and oscillation modes.
The main characteristics of LLs are clearly reflected in the magneto-optical selection rules which cover the dominating ${\Delta\,n=|n^v-n^c|=0}$, the coexistent ${\Delta\,n=0}$ $\&$ ${\Delta\,n=1}$, and the specific ${\Delta\,n=1}$.
These rules for inter-LLs excitations come from the non-equivalence or equivalence of the A$_i$ and B$_i$ sublattices in a supercell.

\end{abstract}
\pacs{PACS:}
\maketitle

\section{Introduction}
Magnetic quantization is one of the mainstream topics in the physical science, such as the rich magneto-electronic properties \cite{me1, me2, me3}, magneto-optical selection rules,\cite{mo1, mo2, mo3} and quantum Hall effects in few-layer graphene systems.\cite{qhe1, qhe2, qhe3}
Diverse physical phenomena could be achieved by changing the atomic components,\cite{at1} the lattice symmetries,\cite{sym1, sym2} the lattice geometries such as planar, buckling, rippled, and folding structures,\cite{geo1, geo2, geo3} the stacking configurations,\cite{conf1, sts3} the number of layers,\cite{lay1, lay2} the distinct dimensionalities,\cite{dim1, dim2} the spin-orbital couplings,\cite{me2, soc1} the single- or multi-orbital hybridizations,\cite{orb1} the electric field,\cite{elec} and the uniform or non-uniform magnetic field.\cite{me2, mag1}
In this Letter, we aim to investigate the interesting quantization phenomena of monolayer graphene under the effect of Si-doped defect.\\

Monolayer graphene presents the unusual essential properties, mainly owing to the hexagonal symmetry and the single-atom thickness.
The isotropic Dirac-cone structures, initiated from the K and K$^\prime$ valleys (corners of the first Brillouin zone), are magnetically quantized into the unique LLs, with the specific energy spectrum proportional to the square root of the magnetic-field strength and quantum number of valence and conduction LLs, $\sqrt{B_zn^{c,v}}$.
This simple relation has been verified by the scanning tunneling spectroscopy (STS),\cite{sts1} optical spectroscopies,\cite{op-sp1} and transport equipment.\cite{qhe1}
The magneto-optical absorption peaks are identified to satisfy a specific selection rule ${\Delta\,n=|n^v-n^c|=1}$, directly reflecting the equivalence of A and B sublattices. Such rule determines the available scattering processes, leading to the unconventional half-integer Hall conductivity of ${\sigma_{xy}}$ = (m + 1/2)4e$^2$/h,\cite{qhe1} in which m is an integer and the factor of 4 represents the spin- and sublattice-dependent degeneracy.
This unusual magnetic quantization is attributed to the quantum anomaly of ${n^{c,v}}$ = 0 LLs associated with the Dirac point.\\

The fundamental properties are efficiently modified by creating a defect effect such as substituted impurities or guest atoms in a hexagonal carbon lattice.
Various guest-atom-doped graphene systems are expected to present the unusual physical phenomena and possess potential applications.
Up to now, carbon host atoms are successfully substituted by the guest atoms of Si,\cite{df1} B,\cite{df2} and N \cite{df2, df3} through the chemical vapor deposition (CVD) or arc discharge methods.
These new 2D materials exhibit the non-equivalence of the original A and B sublattices, leading to the possible existences of energy-gap engineering and the tilted Dirac cone.
According to the first-principles calculations on the Si-doped graphene,\cite{1st-p1, 1st-p2} the $\pi$ bonding extending on a hexagonal lattice is distorted or even destroyed by the different ionization potentials and the non-uniform hopping integrals.
That is, there exists a greatly modified Dirac cone or a significant energy gap, and the Si- and C-dominated low-lying band structure. The drastic changes in energy dispersions, band gap, and atom-dominated wave functions will play critical roles in diversifying the magnetic quantization phenomena.\\

Tight-binding model is an appropriate method to investigate essential magnetic properties of any 2D layered materials, including the magneto-electronic properties,\cite{tbm} magneto-optical and quantum transport properties via the dynamic and static Kubo formulas; magneto-Coulomb excitations within the modified random-phase approximation.
Here, we develop the generalized tight-binding model built from the subenvelope functions on the distinct sublattices, collaborated with the dynamic Kubo formula from linear response theory, to fully explore the diversified electronic and optical properties in Si-doped graphene.
The complex combined effects, which arise from the distinct ionization potentials, the non-uniform hopping integrals $\&$ bond lengths on a deformed hexagonal lattice, the various $B_z$-induced Peierles phases, and the excitations of electromagnetic waves, are accurately included in the huge Hamiltonian matrix.
To overcome such problem in numerical calculations, the exact diagonalization method is proposed to solve magneto-electronic properties and magneto-optical spectra more efficiently.\cite{tbm}
Various kinds of LLs appearing during the variation of Si-distribution configuration and concentration is thoroughly investigated.
Their main features, characterized by probability distributions and oscillation modes, are clearly illustrated by the distinct magneto-optical selection rules.
Apparently, this work could open a new research category in the fundamental properties of 2D layered materials.
The theoretical predictions require further experimental verifications using STS,\cite{sts1, sts2, sts3, sts4} magneto-optical spectroscopies,\cite{op-sp1, op-sp2, op-sp3, op-sp4, op-sp5} and quantum transport measurements.\cite{qhe1} \\

\section{Method}

Four types of typical Si-doped graphene systems can clearly illustrate the diversified properties.
They cover (type I) 2:16 concentration under the Si-(A$_1$, A$_6$)-sublattice distribution (red balls in Fig. 1(a)), (type II) 2:16 concentration for the -(A$_6$, B$_4$) configuration (green balls), (type III) 2:64 concentration related to the ($A_1$, A$_{19}$) sublattices (Fig. 1(b)) , and (type IV) a pristine one with the equivalent A and B sublattices.
The types I and III (type II) presents the non-equivalent (equivalent) A$_i$ and B$_i$ sublattices in a Si-induced unit cell, while both A and B sublattices are fully equivalent for pristine graphene.
For example, the type I has a rectangular traditional cell comprising two Si and fourteen C atoms, which is consistent with the Landau gauge under ${B_z\hat z}$.
There exist a slight buckling near the guest atoms ($\sim$ 0.93 $\AA$ deviation from graphene plane) and the distinct C-C and  Si-C bond length (1.42 $\AA$ $\&$ 1.70 $\AA$), according to the first-principles calculations.\cite{1st-p1, 1st-p2}
Though this indicates remarkable modifications of the $\pi$ bonding extending on a hexagonal lattice, the non-uniform site energies and nearest-neighboring hopping integrals due to the major ${2p_z}$ orbitals of C host atoms and the minor ${3p_z}$ orbitals of Si guest atoms are sufficient in understanding the low-lying energy bands.
These parameters are optimized as ${\epsilon_{Si-C}}$=1.3 eV, ${\gamma_{C-C}}$=2.7 eV and ${\gamma_{Si-C}}$=1.3 eV, respectively, in order to reproduce the band structures from the first-principles calculations.
They are valid for many different distribution configurations and concentrations of Si-doped graphene systems. \\

\begin{figure}[h]
\centering
{\includegraphics[height=12cm]{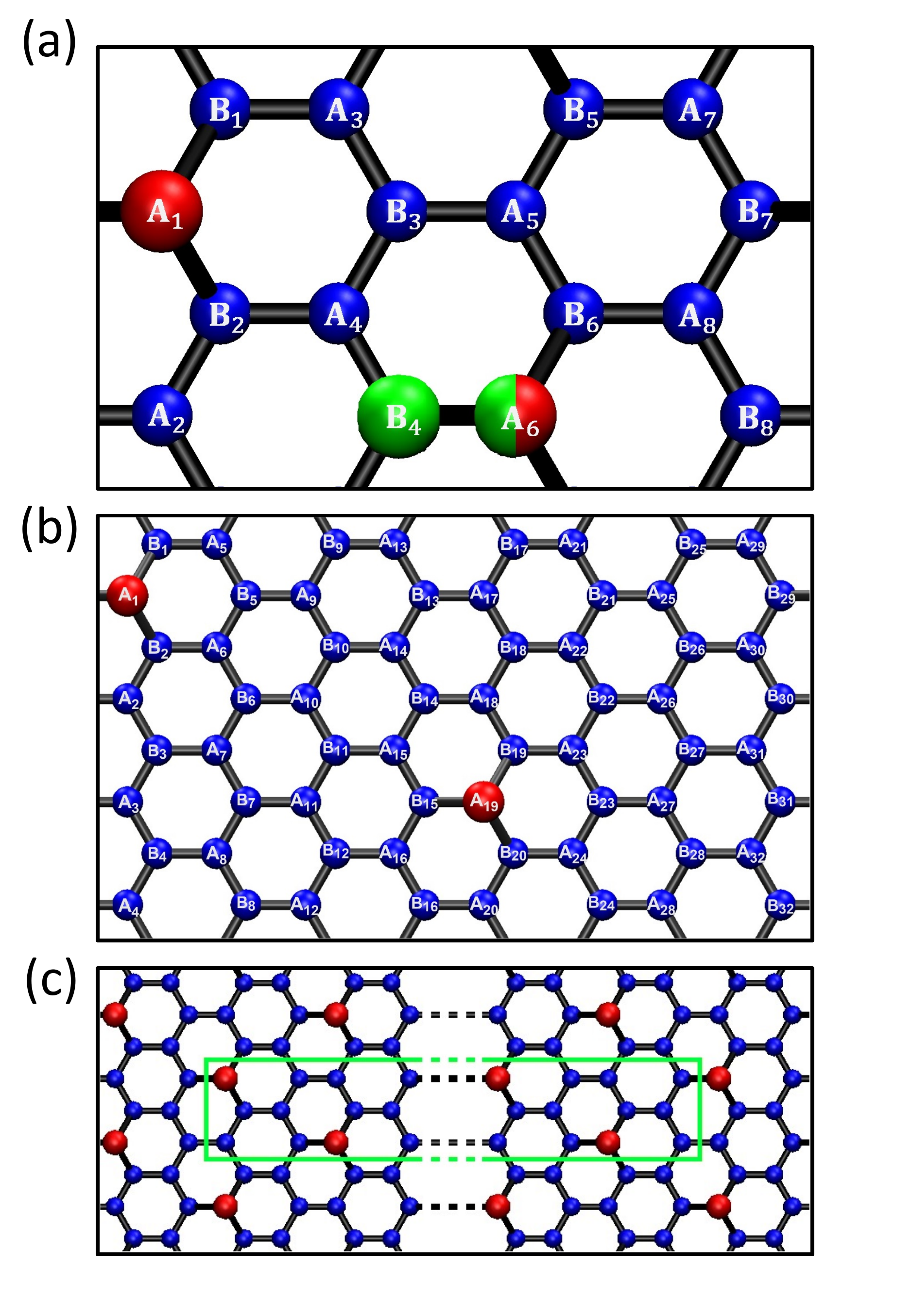}}
\caption{(Color online) Geometric structures of Si-doped graphene systems of (a) (type I) 2:16 concentration under the A$_i$- (red balls) and (type II) [A$_i$, B$_j$]-sublattice distributions (green balls), (b) (type III) 2:64 concentration for the A$_i$-sublattice distribution. An enlarged rectangular unit cell in ${B_z\hat z}$ is presented in (c).}
\label{Figure 1}
\end{figure}

\subsection{Tight-binding Hamiltonian}

In Si-doped graphene, the unit cell is expanded as ($n\vec{a_1}, n\vec{a_2}$), where $\vec{a_1}$ and $\vec{a_2}$ are the lattice constants of pristine graphene and $n$ is the cell multiplicity of the supercell. The concentration of the Si guest atoms in graphene is defined as 1:2$n^2$, as illustrated in Fig. 2.
The low-energy essential properties are mainly determined by the C-${2p_z}$ and ${Si-3p_z}$ orbitals.
There are 2$n^2$ sublattices in a supercell, including ${A_i}$ and ${B_i}$, as clearly shown in Figs. 2(b) and 2(c).
The zero-field Hermitian Hamiltonian matrix covers the non-uniform bond lengths, site energies and nearest-neighboring hopping integrals, which is expressed as

\begin{eqnarray}
\left\{\begin{array}{ll}
                 H_{i+2nj-2n,i+2nj-n}=H_{i+2nj-n,i+2nj-2n}^*=t_{i+2nj-2n,i+2nj-n}f_1 \\  %
                 H_{i+2nj-2n,m(i)+2nj-n+1}=H_{m(i)+2nj-n+1,i+2nj-2n}^*=t_{i+2nj-2n,m(i)+2nj-n+1}f_2 \\
                 H_{i+2nj-2n,i+2n[m(j)+1]-n}=H_{i+2n[m(j)+1]-n,i+2nj-2n}^*=t_{i+2nj-2n,i+2n[m(j)+1]-n}f_3\\
                 H_{1,1}=\epsilon_{Si-C}.
                \end{array} \right.
\end{eqnarray}
Here,
\begin{eqnarray}
t_{\alpha, \beta}=\left\{\begin{array}{ll}
                 \gamma_{Si-C}, \mathrm{if}\hspace{0.5em} \alpha\hspace{0.5em} \mathrm{or}\hspace{0.5em}  \beta\hspace{0.5em} \mathrm{equal}\hspace{0.5em}1\\  %
                 \gamma_{C-C}, \mathrm{otherwise}.
                \end{array} \right.
\end{eqnarray}
$f_{1,2,3}=e^{\mathrm{i}\vec{k}.\vec{R}_{1,2,3}}$, where $\vec{k}$ is the wave vector and $\vec{R}_{1,2,3}$ are the vectors connecting the nearest-neighbor lattice sites, $m(k)$ is modulo function which defined as $m(k)=k+n-2$ mod $n$, and $i,j$ are the integers ($i,j=1,2,3,...,n$). \\

\begin{figure}[h]
\centering
{\includegraphics[height=12cm]{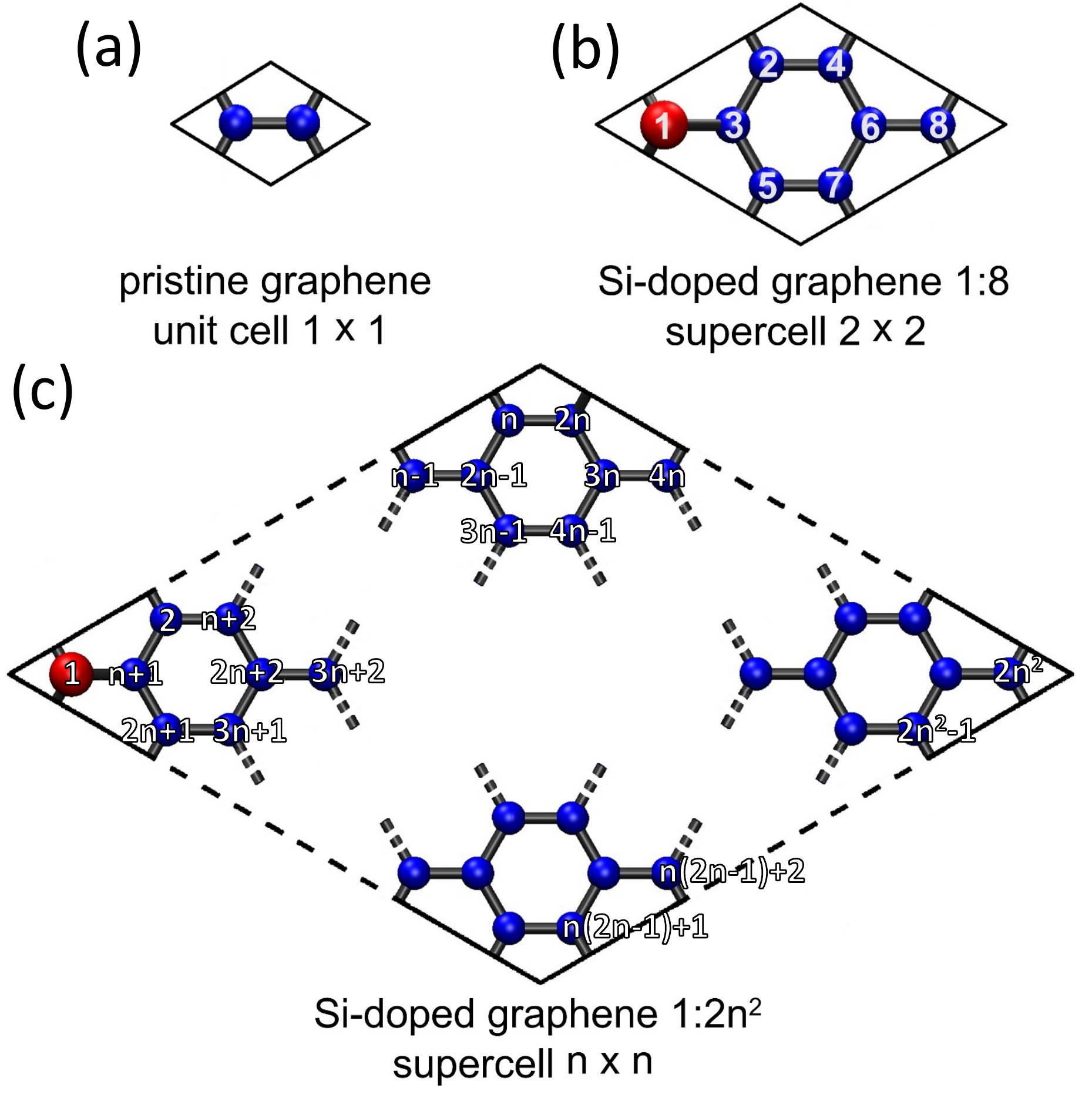}}
\caption{(Color online) The geometric structures of (a) pristine graphene and Si-doped graphene for (b) $n=2$ and (c) arbitrary $n$.}
\label{Figure 2}
\end{figure}

The presence of a uniform magnetic field (${B_z\hat z}$) significantly change the physical features of the systems. 
The dimension of the magnetic Hamiltonian matrix is determined by the guest-atom- and vector-potential-dependent periods, in which the latter is much longer than the former, and their ratio is assumed to be an integer for convenience in calculations.
The vector potential is chosen as ${B_zx\hat y}$ and this creates a position-related Peierls phase of ${\Delta{\bf G}_{mm'}=\frac{2\pi}{\phi_0}\int_{R_{m'}}^{R_m}{\rm\bf A}({\rm\bf r})\cdot {\rm d} {\rm\bf r}}$ in the nearest-neighbor hopping integral.\cite{tbm}
The intrinsic atomic interaction becomes $\gamma_{C-C}\Delta{\bf G}_{mm'}$ and $\gamma_{Si-C}\Delta{\bf G}_{mm'}$.
Due to the periodicity of the Peierls phase, the primitive unit cell is extended in the $\hat{x}$ direction to be a long rectangular, as indicated in Fig. 1(c).
It is noticed that, for convenience, we redefine the original unit cell as a rectangular form, referring to Fig. 1(c).
The magnetic Hamiltonian dimension is changed to become ${4n^2R_B \times 4n^2R_B}$, in which ${R_{B}}$ is defined as the ratio of flux quantum (${\phi_0\,=hc/e}$) versus magnetic flux through each hexagon, e.g., ${R_B}=8000$ at ${B_z=10}$ T.
For the type I of guest-atom distribution, such unit cell covers ${16R_B}$ atoms (${8R_B}$ A and B atoms).
Thus, the Bloch wave functions under a ${B_z\hat z}$ can be expressed in term of the linear superposition of the ${16R_B}$ tight-binding functions in an enlarged unit cell.
The huge complex matrix could be solved more efficiently by transforming it into a band-like one under the rearrangement of the tight-binding function.\cite{tbm}
In addition, the investigation of localization feature of the magnetic wave functions greatly reduces the computation time.\\

After the exact diagonalization of the giant magnetic Hamiltonian, the LL wave function, with quantum number $n^{c,v}$, could be expressed as

\begin{eqnarray}
\Psi (n^{c,v},\mathbf{k})=\sum_{i=1}^{T_{si}}\sum_{\alpha=1}^{R_B} [ A_{i,\alpha} (n^{c,v},\mathbf{k}) |\psi_{i,\alpha} (A) \rangle + B_{i,\alpha} (n,\mathbf{k}) |\psi_{i,\alpha} (B) \rangle ].
\end{eqnarray}

In this notation, $\psi_{i,\alpha}$ is the ${2p_z}$- or ${3p_z}$-orbital  tight-binding function localized at the ${A_i}$ or ${B_i}$ sublattice.
$A_{i,\alpha} (n^{c,v},\mathbf{k}) (B_{i,\alpha} (n^{c,v},\mathbf{k}))$ is the amplitude on the ${A_i}$ (${B_i}$) sublattice.
Specifically, all the amplitudes in an enlarged unit cell could be regarded as the spatial distributions of the sub-envelope functions on the distinct sublattices; they therefore dominate the main features of the LL wave functions.\\

\subsection{Absorption Function and Gradient Approximation}

When the Si-doped graphene exists in an electromagnetic wave, the occupied valence states are vertically excited to the unoccupied conduction ones.
In addition to ${\Delta{\bf k}=0}$, the electric-dipole perturbations require the inter-LL excitations to satisfy a new magneto-optical selection rule of ${\Delta\,n=0}$. Such interesting behavior has never revealed in other layered condensed-matter systems.
According to the linear Kubo formula, the intensity of magneto-optical excitations is characterized by

\begin{eqnarray}
A(\omega) \propto
\sum_{n^c,n^v} \int_{1stBZ} \frac {d\mathbf{k}}{(2\pi)^2}
 \Big| \Big\langle \Psi^{c} (n^c,\mathbf{k})
 \Big| \frac{   \hat{\mathbf{E}}\cdot \mathbf{P}   } {m_e}
 \Big| \Psi^{v}(n^v,\mathbf{k})    \Big\rangle \Big|^2 \nonumber
\end{eqnarray}
\begin{eqnarray}
 \times
Im \Big[      \frac{1}
{E^c (n^c,\mathbf{k})-E^v (n^v,\mathbf{k})-\omega - i\Gamma}           \Big].
\end{eqnarray}

\noindent
The square of velocity matrix element ($\Big\langle \Psi^{c} (n^c,\mathbf{k})  \Big| \frac{   \hat{\mathbf{E}}\cdot \mathbf{P}   } {m_e}
\Big| \Psi^{v}(n^v,\mathbf{k})    \Big\rangle$) determines the available excitation channels and the spectral strength, since it is associated with the spatial distribution symmetries of the initial and final LLs. The second term in the integral function is the delta-function-like joint density of states arising from the inter-LL transitions of  ${(n^v,{\bf k})\rightarrow\,(n^c,{\bf k})}$, in which the broadening factor is ${\Gamma\,=1}$ meV. $\hat{\mathbf{E}}$, $\mathbf{P}$ and $m_e$ are, respectively, unit vector of electric polarization, momentum operator and bare electron mass. Because the direction of the planar electric field hardly affects optical absorption spectra, ${\hat{\mathbf{E}}\parallel\,\hat{\mathbf{x}}}$ is chosen in the current work. \\

The velocity matrix element is evaluated from

\begin{eqnarray}
{ \Big\langle \Psi^{c} (n^c,\mathbf{k}) \Big| \frac{   p_x   } {m_e} \Big| \Psi^{v}(n^v,\mathbf{k})    \Big\rangle} \hspace{20em} \nonumber \\
=\sum_{i,i=1}^{T_{si}}\sum_{\alpha,\alpha'=1}^{R_B}
\{ A_{i',\alpha'}^*(n^v,\mathbf{k})\times A_{i,\alpha}(n^c,\mathbf{k}) \Big\langle \psi_{i',\alpha'} (A)  \Big| \frac{   p_x   } {m_e} \Big| \psi_{i,\alpha} (A)     \Big\rangle  \hspace{0.5em} \nonumber \\
+A_{i',\alpha'}^*(n^v,\mathbf{k})\times B_{i,\alpha}(n^c,\mathbf{k}) \Big\langle \psi_{i',\alpha'} (A)  \Big| \frac{   p_x   } {m_e} \Big| \psi_{i,\alpha} (B)     \Big\rangle \hspace{0.5em} \nonumber \\
+B_{i',\alpha'}^*(n^v,\mathbf{k})\times A_{i,\alpha}(n^c,\mathbf{k}) \Big\langle \psi_{i',\alpha'} (B)  \Big| \frac{   p_x   } {m_e} \Big| \psi_{i,\alpha} (A)     \Big\rangle  \hspace{0.5em} \nonumber \\
+B_{i',\alpha'}^*(n^v,\mathbf{k})\times B_{i,\alpha}(n^c,\mathbf{k}) \Big\langle \psi_{i',\alpha'} (B)  \Big| \frac{   p_x   } {m_e} \Big| \psi_{i,\alpha} (B)     \Big\rangle \}.
\end{eqnarray}

The critical dipole factor is evaluated from the gradient approximation, as successfully utilized in carbon-related ${sp^2}$-bonding systems.\cite{approx} That is,

\begin{eqnarray}
\Big\langle \Psi^{c} (n^c,\mathbf{k})
 \Big| \frac{   p_x   } {m_e}
 \Big| \Psi^{v}(n^v,\mathbf{k})    \Big\rangle
\cong \frac{\partial}{\partial k_x}\Big\langle \Psi^{c} (n^c,\mathbf{k})
 \Big| H
 \Big| \Psi^{v}(n^v,\mathbf{k})    \Big\rangle.
\end{eqnarray}

\noindent
Equation (6) clearly indicates that the electric-dipole magneto-optical excitations are dominated by the A$_i$ (B$_i$) subenvelope functions of the initial ${n^v}$ LL and the B$_{i^{\prime}}$ (A$_{i^{\prime}}$) ones of the final ${n^c}$ LL. Since the velocity matrix element is associated with the $\bf{k}$-dependent nearest-neighbor hopping integrals, $i$ and $i^{\prime}$ denote the nearest-neighbor lattice sites.
By accurate calculations and detailed examinations on the well-behaved LLs, only the first and second terms in Eq. (6) make the significant contributions.
Most importantly, the available transition channels need to satisfy the ${\Delta\,n\,= 0}$ selection rule so that the quantum mode of the initial valence LL state at the ${A_i}$ (${B_i}$) sublattice is identical to that of the final conduction LL state at the ${B_i}$ (${A_i}$) sublattice. \\ 

\section{Results and Discussion }

\subsection{Electronic structure}
The Si-doped graphene exhibits the unusual low-energy electronic properties.
For the type I (the red curves in Fig. 3(a)), the valence and conduction bands, nearest to the Fermi level ($E_F$), have the parabolic energy dispersions separated by a direct energy gap of ${E_g=0.74}$ eV.
The electronic energy spectrum is anisotropic along the different ${\bf k}$-directions, and it is asymmetric about $E_F$.
Similar results are also revealed in the type III of lower-concentration system with a 0.26 eV band gap (the blue curves in Fig. 3(a)).
Energy gap appears only if the guest atoms are situated at either the A$_i$ or B$_i$ sublattices.
The non-uniform site energies and hopping integrals further induce the partial termination of the $\pi$ bonding (the minor localized states), as observed in the zero-field and magnetic wave functions (Figs. 4(a), 4(b); 4(d)).
On the other side, $E_g$ vanishes for the type II distribution configuration (the solid curve in Fig. 2(b)).
The guest-atom distribution with equal weight induces the distorted $\pi$ and thus the strongly modified Dirac cone structure with an obvious shift of Dirac point, the reduced Fermi velocity, and the anisotropic energy spectrum.
Apparently, graphene exhibits a well-behaved Dirac cone (the dashed curve) because of the purely hexagonal symmetry.\\

\begin{figure}[h]
\centering
{\includegraphics[height=7cm]{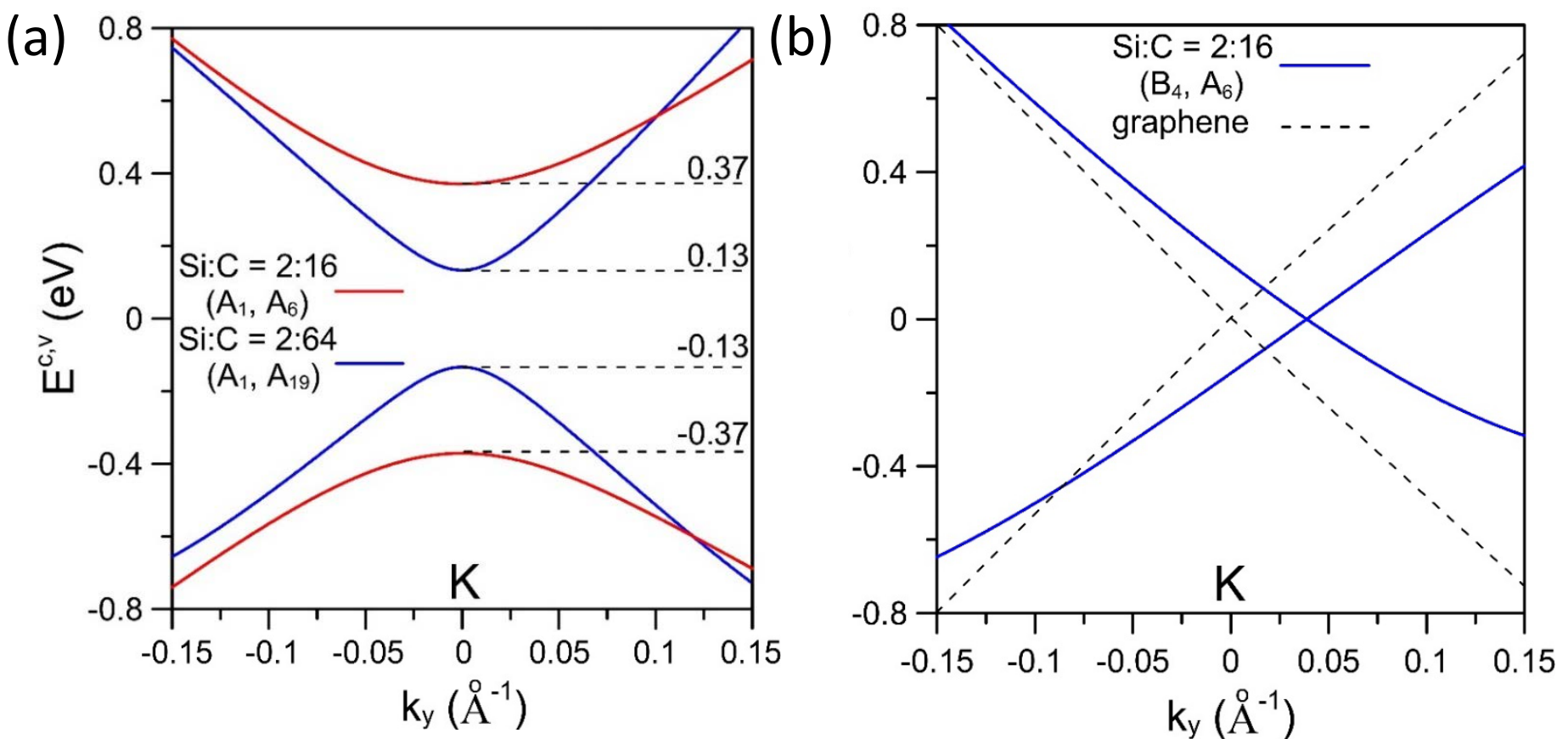}}
\caption{(Color online) The (a)-(c) low-lying energy bands for three types of Si distributions and concentrations as mentioned in Fig. 1. The Dirac cone of pristine graphene (type IV) is also shown in Fig. 2(b) for comparison.}
\label{Figure 3}
\end{figure}

\subsection{The quantized Landau levels}

The magneto-electronic properties exhibit the rich and unique features.
The low-lying LL energy spacings, as shown at ${B_z=10}$ T in Figs. 4(a) and 4(b), are almost uniform and have an energy gap close to the zero-field value.
In general, the quantum number of each LL is defined from the zero points of the dominating oscillation mode.
For the ${12.5}$ $\%$ Si-A$_i$-sublattice graphene, the magnetic Bloch wave function arises from the subenvelope functions of the 16 tight-binding functions on the corresponding sublattices.
Its spatial probability distribution of the ${(k_x=0, k_y=0)}$ state is localized at (1/6 $\&$ 4/6) and (2/6 $\&$ 5/6) of an enlarged unit cell (Fig. 1(c)).
Any ${(k_x,k_y)}$ LL states in the reduced first Brillouin are doubly degenerate except for the spin degree of freedom.
Apparently, the decoration of Si guest atoms leads to the destruction of the planar inversion symmetry and thus the non-degenerate 1/6 and 2/6 LL states.
According to the neighboring chemical environment, the original 16 sublattices could be classified into four subgroups of (A$_1$, A$_6$), (A$_2$, A$_3$, A$_4$, A$_5$, A$_7$, A$_8$), (B$_1$, B$_2$, B$_4$, B$_5$, B$_6$, B$_7$), and (B$_3$, B$_8$).
The low-energy conduction LL states are dominated by the A$_1$ sublattice with the Si-${3p_z}$ tight-binding function, so that the zero-point number of the well-behaved probability distribution could serve as a good quantum number.
${n^c=}$0, 1, 2 and so on appears in the normal sequence. Specifically, the contributions from the B$_3$ sublattice are small, as seen in the zero-field wave functions.
The oscillation modes are characterized by $n^c$ for the significant sublattices except for the weak ${n^c\pm\,1}$ B$_3$ sublattice.
On the other hand, the valence LL states mainly originate from all the B$_i$ sublattices of the C-${2p_z}$ tight-binding functions, where they have the similar oscillation modes in determining $n^v$.
The contributions from the A$_i$ sublattices are very small, and the number of zero points is ${n^v-1}$ or ${n^v+1}$ (Figs. 4(a) and 4(b)).
The sequences of ${n^c}$ and ${n^v}$ present good orderings, i.e., the crossing or anti-crossing behaviors are absent.
These reliable magneto-electronic properties are very useful in understanding the rich magneto-optical excitation spectra.\\

\begin{figure}[h]
\centering
{\includegraphics[height=9cm]{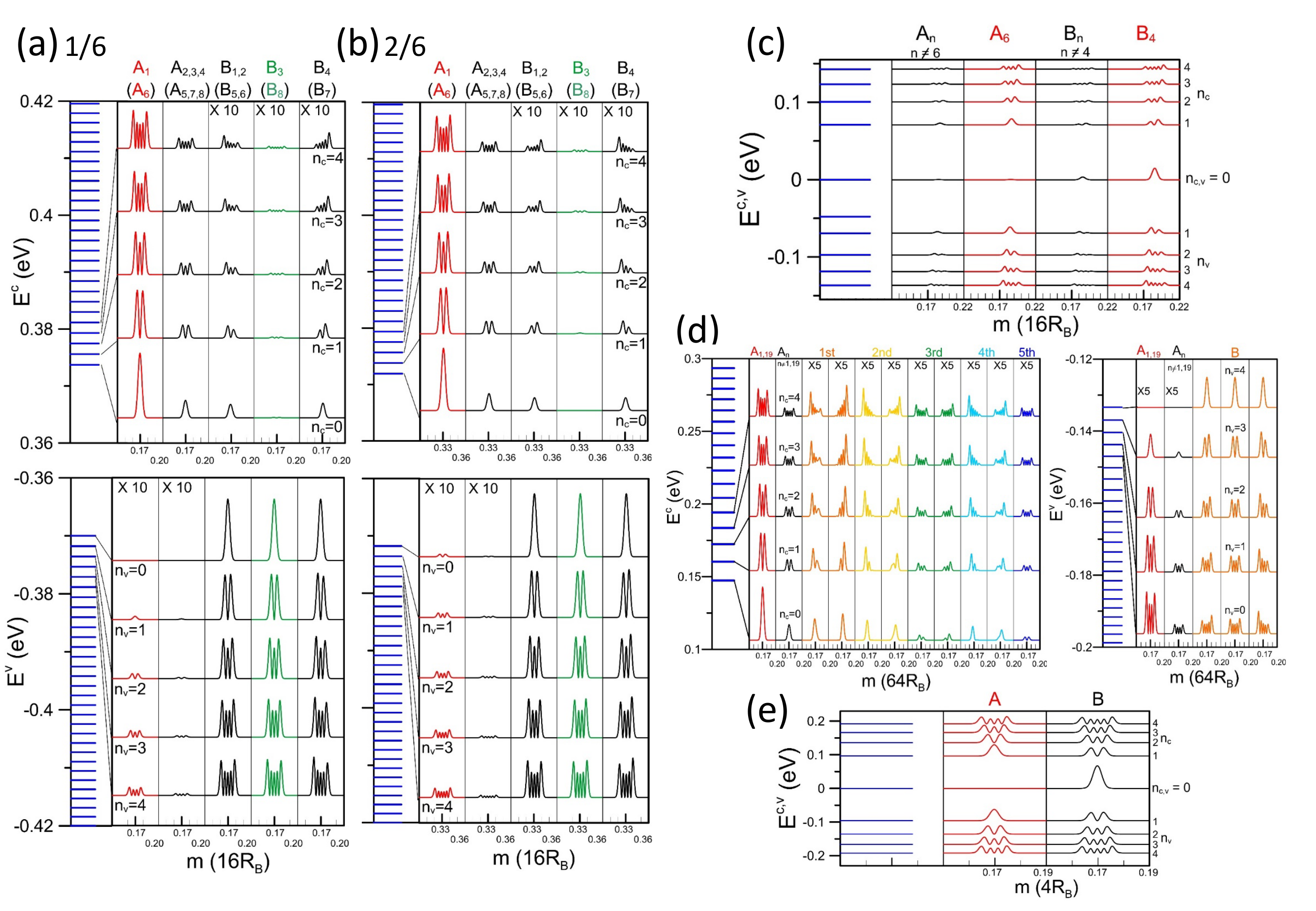}}
\caption{(Color online)  The conduction and valence LL energy spectra and corresponding probability distributions for (a)-(b) type I near the 1/6  and 2/6 localization centers at ${B_z=10}$ T. Similar plots for (c) type II, (d) type III and (e) type IV are also presented. In Fig. (d), $B_{1st}$ = $\{B_{1, 2, 15, 19, 20, 29}\}$, $B_{2nd}$ = $\{B_{5, 14, 16, 23, 30, 32}\}$, $B_{3rd}$ = $\{B_{3, 4, 6, 8, 11, 12, 17, 18, 22, 24, 25, 26}\}$, $B_{4th}$ = $\{B_{9, 10, 13, 27,28, 31}\}$, and $B_{5th}$ = $\{B_{7, 21}\}$ denote the B sublattices which are from the nearest to the furthest to the doped Si atoms, respectively.}
\label{Figure 4}
\end{figure}

The spatial oscillation modes are very sensitive to the changes in the distribution configuration and concentration of guest atoms.
There exist four kinds of LLs, corresponding to four types of lattice geometries.
For a very strong non-equivalence between A$_i$ and B$_i$ sublattices and enough high concentration (2:16 under the Si-(A$_1$, A$_6$) configuration in Fig. 1(a)), only the significant sublattices exhibit the similar oscillation modes for the low-lying valence and conduction LLs (the first kind in Figs. 4(a)-4(b)).
However, the enhanced equivalence (green balls in Fig. 1(a)) and the reduced concentration (Fig. 1(b)) can create the composite behaviors related to the heavily non-equivalent A$_i$ $\&$ B$_i$ sublattices and the fully equivalent ones (e.g., pristine graphene).
The former, with two Si atoms in A$_6$ and B$_4$ sublattices, has the highly equivalent environment.
All the sublattices make significant contributions to the LL wave functions, in which the difference of zero point number is ${\pm\,1}$ for  A$_i$ and B$_i$ sublattices (the second kind in Fig. 4(c)).
Specifically, their spatial distributions are highly asymmetric and localization centers seriously deviate from 1/6 $\&$ 2/6, directly reflecting the seriously titled Dirac-cone (Fig. 3(b)).
Also, a seriously distorted distribution consists of the main ${n^{c,v}}$ mode and  the side ${n^{c,v}\pm\,1}$ ones.
The localization centers are recovered to the normal positions under the decrease of concentration with the Si-A$_i$ distribution (2:64 in Fig. 1(b)).
The certain B$_i$ sublattices, which are farthest from the Si atom and possess ${n^c\pm\,1}$ modes, become observable for the conduction LLs, and so do for the A$_i$ sublattices in valence LLs (the third kind in Fig. 4(d)).
Moreover, the wave functions in other B$_i$ sublattices presents the highly asymmetric distributions for the Si-dominated LLs.
Finally, a pristine graphene displays the well-behaved LLs about the localization centers and the difference of ${\pm\,1}$ in the zero-point number due to the equivalent A and B sublattices (the fourth kind in Fig. 4(e)).\\

\begin{figure}[h]
\centering
{\includegraphics[height=9cm]{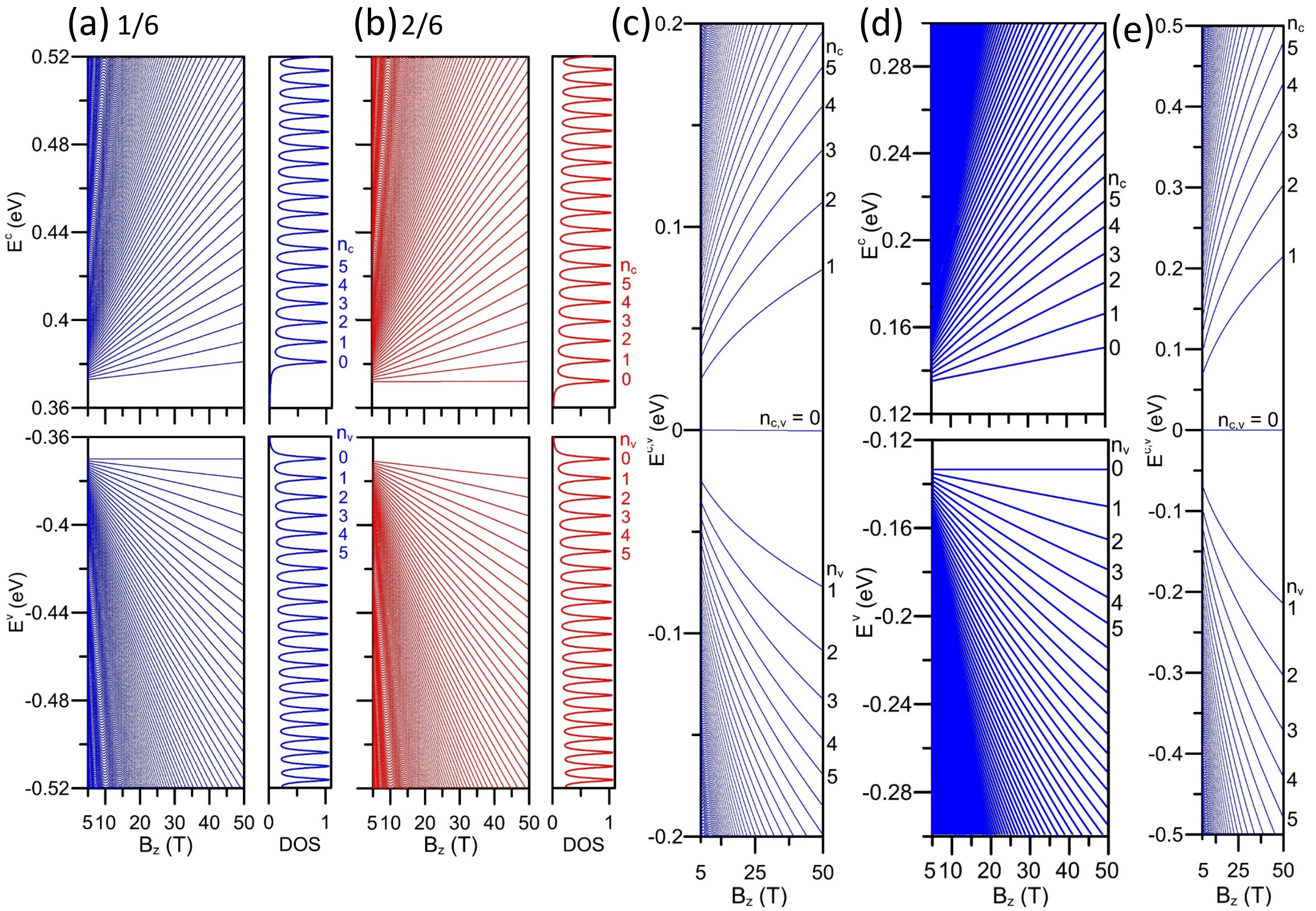}}
\caption{(Color online)  The (a)-(e) $B_z$-dependent LL energy spectra corresponding to four types of lattice geometries in Fig. 3. The density of states are also shown for the type I of Si distribution configuration.}
\label{Figure 5}
\end{figure}

The $B_z$-dependent LL energy spectrum, as clearly indicated in Figs. 5(a)-5(e), presents the unusual features.
The crossing or anti-crossing behaviors are forbidden for the low-lying LLs, illustrating the well separated LL states and the specific-mode wave functions.
For the first and third kinds of LLs (Figs. 5(a), 5(b) and 5(d)), the dispersion relation is almost linear, and the LL energy spacing is uniform.
Specifically, the initial valence and conduction LLs, which are, respectively, related to the 1/6 and 2/6 localization centers, remain the fixed energies during the variation of field strength.
They purely come from the localized electronic states, since the magnetic wave functions vanish in all the A$_i$ or the B$_i$ sublattices, as observed from Figs. 4(a) and 4(b).
That is, the termination of the $\pi$ bonding appears on a guest-host mixed hexagonal lattice.
A uniform perpendicular magnetic field can create the splitting of the localized and extended electronic states; otherwise, they are hybridized each other and are revealed near the K and K$^\prime$ valleys.
Such LL states could be examined from the STS measurements on the van Hove singularities of the density of states, e.g., the delta-function-like prominent peaks across the Fermi level (Figs. 5(a) and 5(b)).
On the other side, the second and fourth kinds of LLs shows the ${\sqrt B_z}$-dependent energy spectra except for the constant energy of the degenerate ${n^{c,v}=0}$ LLs.
Apparently, the latter has the largest energy spacing among four kinds of  LLs because of the lowest density of states.\\

\subsection{The magneto-optical selection rules}

The main features of LLs are directly reflected in magneto-optical absorption spectra with a lot of delta-function-like peaks, as demonstrated in Fig. 6.
For the Si-A$_1$-doped graphene with 2:16 concentration (Fig. 6(a)), the spectral intensity gradually declines with the increasing frequency, while the energy spacing between two neighboring absorption peaks is almost uniform.
Only the inter-LL transitions, which correspond to the identical quantum mode in the valence and conduction LLs, are revealed as the significant absorption peaks.
For example, the threshold frequency due to ${0^v\rightarrow\,0^c}$ is 0.743 eV very close to the energy gap.
The magneto-optical selection rule, ${\Delta\,n=0}$, could be thoroughly examined from the electric-dipole momentum in Eq. (6).
It is mainly dominated by the specific Hamiltonian matrix elements covering the nearest-neighboring hopping integrals.
As a result, the effective vertical excitations depend on the subenvelope functions of the B$_i$/A$_i$ $\&$ B$_{i+1}$/A$_{i}$ sublattices in the $n^v$/$n^c$ LLs.
Furthermore, the significant sublattices present the same zero-point number. These results are responsible for a new selection rule, is never found in other condensed-matter systems up to now.\\

\begin{figure}[h]
\centering
{\includegraphics[height=12cm]{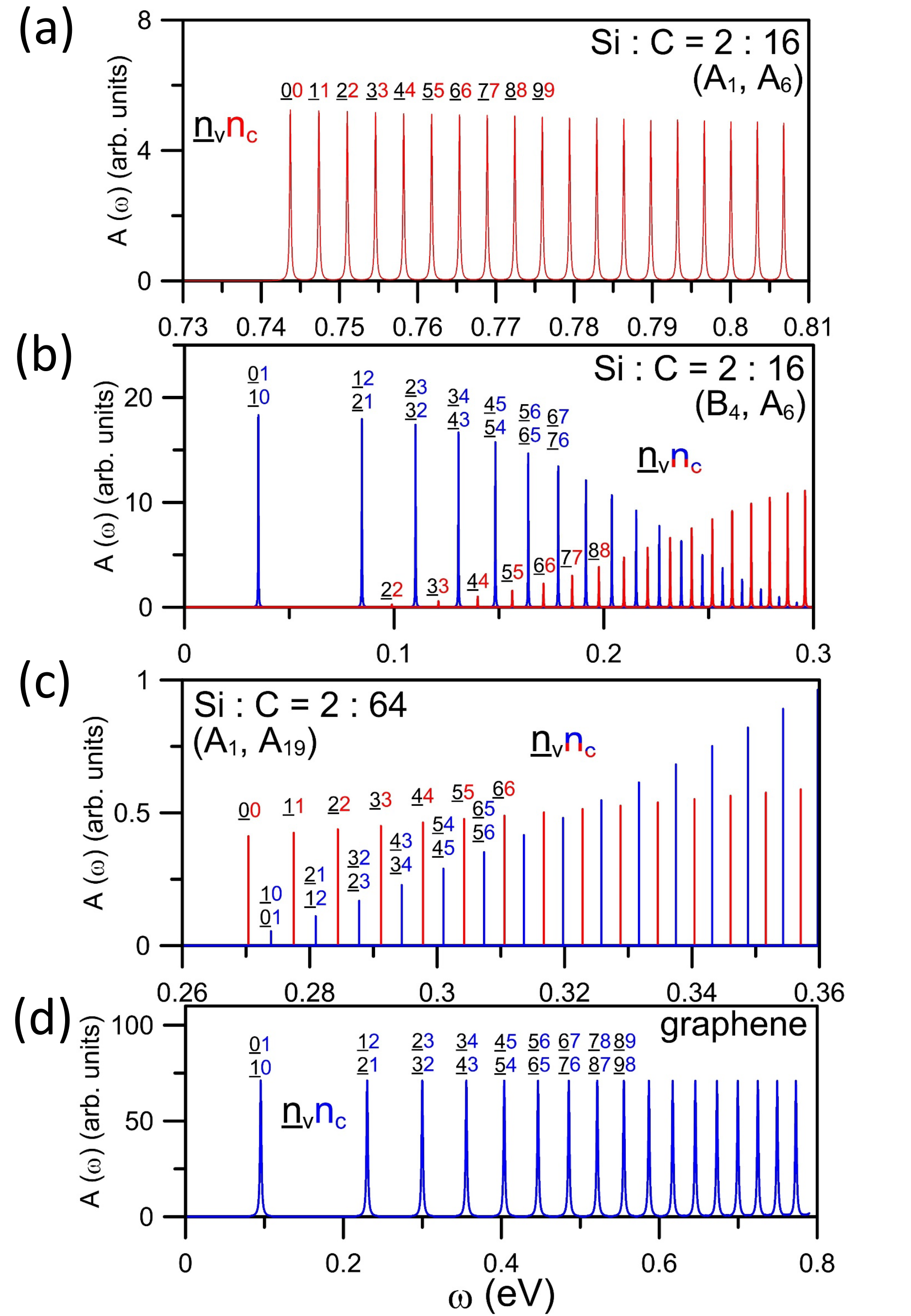}}
\caption{(Color online) The magneto-optical absorption spectra of (a) type I, (b) type II, (c) type III, and (d) type IV of Si-doped graphene systems.}
\label{Figure 6}
\end{figure}

Both ${\Delta\,n=0}$ and 1 come to exist together under the reduced non-equivalence of A$-i$ and B$_i$ sublattices, as clearly shown in Figs. 6(b) and 6(c).
As to the Si-(A$_3$, B$_5$)-decorated graphene of 2:16 concentration, two categories of inter-LL excitation channels frequently appear during the variation of frequency.
The absorption peaks of ${\Delta\,n=1}$ decrease quickly, while the opposite is true for those of ${\Delta\,n=0}$. The former and the latter, respectively, come from the neighboring A$_i$ and B$_i$ sublattices with the mode difference of ${\pm\,1}$ and 0.
The lower-frequency absorption peaks are dominated by ${\Delta\,n=1}$, since the corresponding LLs, being similar to those of graphene, are magnetically quantized from the low-lying titled Dirac cone.
However, with the increasing energy, the enlarged derivation of localization center and the enhanced distortion of spatial probability (Fig. 4(c)) create and enhance the available channels of ${\Delta\,n=0}$ through the strengthened side mode.
Such characteristics lead to the strong competition between these two kinds of magneto-selection rules.\\

On the other hand, the coexistent selection rules present another kind of behavior for the Si-A$_i$-doped graphene with the reduced concentration (Fig. 6(c)).
The ${\Delta\,n=0}$ channels dominate the lower-frequency absorption spectrum, since the significant sublattices possess the same oscillation modes (Fig. 4(d)).
Their peak intensities slowly grow with the increasing frequency, clearly indicating the significant competition or cooperation between two categories of inter-sublattice transitions. ${B_{i}\rightarrow\,A_{i}}$ and ${A_{i}\rightarrow\,B_{i}}$ (except for the furthest ones) appear under the ${n^v\rightarrow\,n^c}$  inter-LL transition, in which the second category is absent for a sufficiently high concentration in Fig. 6(a).
Especially, the quick enlargement of the ${\Delta\,n=1}$ absorption peaks is due to the enhanced oscillations in the A$_i$ sublattices of valence LLs $\&$ the furthest $B_i$ sublattices of conduction LLs, and the strengthened side modes in other $B_i$ sublattices.
On the other hand, it is well known that graphene only presents the ${\Delta\,n=1}$ absorption peaks with a uniform optical spectrum (Fig. 6(d)) as a result of the full equivalence of A and B sublattices.
Among all the Si-doped graphene systems, the pristine one has the strongest intensity and the largest energy spacings.
These are closely related to the smallest density of states of the isotropic Dirac cone, as indicated from a detailed comparison with the separated parabolic bands and the titled Dirac cone (Fig. 3).\\

The diverse magneto-electronic properties and absorption spectra could be verified by STS and optical spectroscopies, respectively.
STS is a very efficient method for examining the quantized energy spectra.
The tunneling differential conductance (dI/dV) is approximately proportional to the DOS, and it directly reflects the structure, energy, number and height of the LL peaks.
Part of the theoretical predictions have been confirmed under the magnetic measurements, such as the ${\sqrt B_z}$-dependent LL energies in monolayer graphene,\cite{sts1} the linear $B_z$-dependence in bilayer AB stacking,\cite{sts2} the coexistence of the square-root and linear  $B_z$-dependences in trilayer AB-stacked graphene,\cite{sts3} and the 3D and 2D characteristics of the Landau subbands in Bernal graphite.\cite{sts4}
Magnetic quantization phenomena of layered systems could also be identified by magneto-optical spectroscopies.\cite{op-sp1, op-sp2, op-sp3, op-sp4, op-sp5}
The examined phenomena are exclusive in graphene-related systems, such as 0D LLs in few-layer graphenes and 1D Landau subbands in bulk graphites.\cite{op-sp1, op-sp2, op-sp3}
A lot of prominent delta-function-like absorption peaks are clearly shown by the inter-LL excitations due to massless and massive Dirac fermions in monolayer \cite{op-sp1} and AB-stacked bilayer graphenes.\cite{op-sp2}
The former and the latter absorption frequencies are square-root and linearly proportional to $B_z$, respectively.
As to the inter-Landau-subband excitations in Bernal graphite, one could observe a strong dependence on the wave vector $k_z$, which characterizes both kinds of Dirac quasi-particles.\cite{op-sp4, op-sp5}
In short, the experimental examinations on four kinds of LLs and the distinct magneto-optical selection rules could provide the full information about the diversified essential properties, establish the emergent binary or ternary graphene compounds, and confirm the developed theoretical framework.\\

\section{Concluding Remarks}

The Si-doped graphene systems, the emergent 2D binary compound materials, are worthy of the systematically theoretical and experimental researches and very suitable for exploring the novel physical phenomena.
These systems have revealed the diverse electronic and optical properties under the magnetic quantization, being absent in other condensed-matter systems.
There exist four kinds of LLs, according to the probability distributions and oscillation modes on the distinct sublattices, and the relations between A$_i$ and B$_i$ sublattices.
They cover the significant B$_i$ sublattices of valence LLs $\&$ $A_i$ sublattices of conduction LLs with the same modes, the observable (A$_i$, B$_i$) sublattices with a mode difference of ${\pm\,1}$, the serious deviations of localization centers $\&$ the highly asymmetric distributions composed of the main and side modes, the same modes for valence B$_i$ and conduction A$_i$ sublattices, the ${\pm\,1}$ zero-point differences between valence A$_i$  and conduction B$_i$ sublattices $\&$ the perturbed multi-modes in most of conduction B$_i$ sublattices (except for the furthest ones); the oscillator-like oscillation modes with the equivalent A and B sublattices.
Such LLs lead to the unusual magneto-optical selection rules of the dominating ${\Delta\,n=0}$, the coexistent ${\Delta\,n=1}$ $\&$ 0 with strong competitions, and the specific ${\Delta\,n=1}$.
The interesting features of LLs correspond to various concentrations and distribution configurations; they are, the Si-A$_i$-doped graphene with an enough high concentration, the (A$_i$, B$_i$)-decorated graphene, the low-concentration A$_i$-doped system, and the pristine one.
In this work, the generalized theoretical framework takes into account simultaneously all the critical factors of non-uniform bond lengths, site energies and hopping integrals, and external field effects without the perturbation forms.
This method is expected to be very useful in fully understanding the essential properties of the main-stream layered systems.  \\

\begin{acknowledgements}
This material is based upon work supported by the Air Force Office of Scientific Research (AFOSR) under award number FA2386-18-1-0120. D.H. thanks the supports from the AFOSR and from the DoD Lab-University Collaborative Initiative (LUCI) Program.
\end{acknowledgements}

\newpage

$\textbf{References}$


\begin{references}
\bibitem{me1} Yin, L.-J.; Bai, K.-K.; Wang, W.-X.; Li, S.-Y.; Zhang, Y.; He, L. Landau Quantization of Dirac Fermions in Graphene and Its Multilayers. Front. Phys. 2017, 12 (4), 127208.

\bibitem{me2} Do, T.-N.; Shih, P.-H.; Gumbs, G.; Huang, D.; Chiu, C.-W.; Lin, M.-F. Diverse Magnetic Quantization in Bilayer Silicene. Phys. Rev. B 2018, 97 (12), 125416.

\bibitem{me3} Lado, J. L.; Fernández-Rossier, J. Landau Levels in 2D Materials Using Wannier Hamiltonians Obtained by First Principles. 2D Mater. 2016, 3 (3), 035023.

\bibitem{mo1} Stepniewski, R.; Pastor, K.; Grynberg, M. Gamma 8 -Bands Anisotropy and Selection Rules for Far-Infrared Magneto-Optical Transitions in HgTe. J. Phys. C: Solid State Phys. 1980, 13 (31), 5783.

\bibitem{mo2} Gopalan, S.; Furdyna, J. K.; Rodriguez, S. Inversion Asymmetry and Magneto-Optical Selection Rules in n-Type Zinc-Blende Semiconductors. Phys. Rev. B 1985, 32 (2), 903–913.

\bibitem{mo3} Wu, J.-Y.; Chen, S.-C.; Do, T.-N.; Su, W.-P.; Gumbs, G. The Diverse Magneto-Optical Selection Rules in Bilayer Black Phosphorus. arXiv:1712.09508 [cond-mat] 2017.

\bibitem{qhe1} Novoselov, K. S.; Geim, A. K.; Morozov, S. V.; Jiang, D.; Katsnelson, M. I.; Grigorieva, I. V.; Dubonos, S. V.; Firsov, A. A. Two-Dimensional Gas of Massless Dirac Fermions in Graphene. Nature 2005, 438 (7065), 197–200.

\bibitem{qhe2} Novoselov, K. S.; McCann, E.; Morozov, S. V.; Fal'ko, V. I.; Katsnelson, M. I.; Zeitler, U.; Jiang, D.; Schedin, F.; Geim, A. K. Unconventional Quantum Hall Effect and Berry's Phase of 2p in Bilayer Graphene. Nature Physics 2006, 2 (3), 177–180.

\bibitem{qhe3} Do, T.-N.; Chang, C.-P.; Shih, P.-H.; Wu, J.-Y.; Lin, M.-F. Stacking-Enriched Magneto-Transport Properties of Few-Layer Graphenes. Phys. Chem. Chem. Phys. 2017, 19 (43), 29525–29533.

\bibitem{at1} John, R.; Merlin, B. Optical Properties of Graphene, Silicene, Germanene, and Stanene from IR to Far UV – A First Principles Study. Journal of Physics and Chemistry of Solids 2017, 110, 307–315.

\bibitem{sym1} Weinberg, M.; Staarmann, C.; Ölschläger, C.; Simonet, J.; Sengstock, K. Breaking Inversion Symmetry in a State-Dependent Honeycomb Lattice: Artificial Graphene with Tunable Band Gap. 2D Mater. 2016, 3 (2), 024005.

\bibitem{sym2} Do, T.-N.; Shih, P.-H.; Chang, C.-P.; Lin, C.-Y.; Lin, M.-F. Rich Magneto-Absorption Spectra of AAB-Stacked Trilayer Graphene. Phys. Chem. Chem. Phys. 2016, 18 (26), 17597–17605.

\bibitem{geo1} Shih, P.-H.; Chiu, C.-W.; Wu, J.-Y.; Do, T.-N.; Lin, M.-F. Coulomb Scattering Rates of Excited States in Monolayer Electron-Doped Germanene. Phys. Rev. B 2018, 97 (19), 195302.

\bibitem{geo2} Lee, K. J.; Kim, D.; Jang, B. C.; Kim, D.-J.; Park, H.; Jung, D. Y.; Hong, W.; Kim, T. K.; Choi, Y.-K.; Choi, S.-Y. Multilayer Graphene with a Rippled Structure as a Spacer for Improving Plasmonic Coupling. Advanced Functional Materials 2016, 26 (28), 5093–5101.

\bibitem{geo3} Liu, F.; Song, S.; Xue, D.; Zhang, H. Folded Structured Graphene Paper for High Performance Electrode Materials. Advanced Materials 2012, 24 (8), 1089–1094.

\bibitem{conf1} Lin, C.-Y.; Do, T.-N.; Huang, Y.-K.; Lin, M.-F. Optical Properties of Graphene in Magnetic and Electric Fields. IOP Publishing 2017, 2053-2563, ISBN: 978-0-7503-1566-1.

\bibitem{sts3} Jhang, S. H.; Craciun, M. F.; Schmidmeier, S.; Tokumitsu, S.; Russo, S.; Yamamoto, M.; Skourski, Y.; Wosnitza, J.; Tarucha, S.; Eroms, J.; et al. Stacking-Order Dependent Transport Properties of Trilayer Graphene. Phys. Rev. B 2011, 84 (16), 161408.

\bibitem{lay1} Pontes, R. B.; Miwa, R. H.; da Silva, A. J. R.; Fazzio, A.; Padilha, J. E. Layer-Dependent Band Alignment of Few Layers of Blue Phosphorus and Their van Der Waals Heterostructures with Graphene. Phys. Rev. B 2018, 97 (23), 235419.

\bibitem{lay2} Zhao, S.; Lv, Y.; Yang, X. Layer-Dependent Nanoscale Electrical Properties of Graphene Studied by Conductive Scanning Probe Microscopy. Nanoscale Res Lett 2011, 6 (1), 498.

\bibitem{dim1} Mostofizadeh, A.; Li, Y.; Song, B.; Huang, Y. Synthesis, Properties, and Applications of Low-Dimensional Carbon-Related Nanomaterials https://www.hindawi.com/journals/jnm/2011/685081/ (accessed Aug 27, 2018).

\bibitem{dim2} Meunier, V.; Souza Filho, A. G.; Barros, E. B.; Dresselhaus, M. S. Physical Properties of Low-Dimensional ${sp}^{2}$-Based Carbon Nanostructures. Rev. Mod. Phys. 2016, 88 (2), 025005.

\bibitem{soc1} Krasovskii, E. E. Spin–Orbit Coupling at Surfaces and 2D Materials. J. Phys.: Condens. Matter 2015, 27 (49), 493001.

\bibitem{orb1} Wang, S. A Comparative First-Principles Study of Orbital Hybridization in Two-Dimensional C, Si, and Ge. Phys. Chem. Chem. Phys. 2011, 13 (25), 11929–11938.

\bibitem{elec} Novoselov, K. S.; Geim, A. K.; Morozov, S. V.; Jiang, D.; Zhang, Y.; Dubonos, S. V.; Grigorieva, I. V.; Firsov, A. A. Electric Field Effect in Atomically Thin Carbon Films. Science 2004, 306 (5696), 666–669.

\bibitem{mag1} Park, S.; Sim, H.-S. Magnetic Edge States in Graphene in Nonuniform Magnetic Fields. Phys. Rev. B 2008, 77 (7), 075433.

\bibitem{op-sp1} Jiang, Z.; Henriksen, E. A.; Tung, L. C.; Wang, Y.-J.; Schwartz, M. E.; Han, M. Y.; Kim, P.; Stormer, H. L. Infrared Spectroscopy of Landau Levels of Graphene. Phys. Rev. Lett. 2007, 98 (19), 197403.

\bibitem{df1} Zhou, W.; Kapetanakis, M. D.; Prange, M. P.; Pantelides, S. T.; Pennycook, S. J.; Idrobo, J.-C. Direct Determination of the Chemical Bonding of Individual Impurities in Graphene. Phys. Rev. Lett. 2012, 109 (20), 206803.

\bibitem{df2} Panchakarla, L. S.; Subrahmanyam, K. S.; Saha, S. K.; Govindaraj, A.; Krishnamurthy, H. R.; Waghmare, U. V.; Rao, C. N. R. Synthesis, Structure, and Properties of Boron- and Nitrogen-Doped Graphene. Advanced Materials 2009, 21 (46), 4726–4730.

\bibitem{df3} Qu, L.; Liu, Y.; Baek, J.-B.; Dai, L. Nitrogen-Doped Graphene as Efficient Metal-Free Electrocatalyst for Oxygen Reduction in Fuel Cells. ACS Nano 2010, 4 (3), 1321–1326.

\bibitem{1st-p1} Zhang, S. J.; Lin, S. S.; Li, X. Q.; Liu, X. Y.; Wu, H. A.; Xu, W. L.; Wang, P.; Wu, Z. Q.; Zhong, H. K.; Xu, Z. J. Opening the Band Gap of Graphene through Silicon Doping for the Improved Performance of Graphene/GaAs Heterojunction Solar Cells. Nanoscale 2015, 8 (1), 226–232.

\bibitem{1st-p2} Shahrokhi, M.; Leonard, C. Tuning the Band Gap and Optical Spectra of Silicon-Doped Graphene: Many-Body Effects and Excitonic States. Journal of Alloys and Compounds 2017, 693, 1185–1196.

\bibitem{sts1} Miller, D. L.; Kubista, K. D.; Rutter, G. M.; Ruan, M.; Heer, W. A. de; First, P. N.; Stroscio, J. A. Observing the Quantization of Zero Mass Carriers in Graphene. Science 2009, 324 (5929), 924–927.

\bibitem{sts2} Rutter, G. M.; Jung, S.; Klimov, N. N.; Newell, D. B.; Zhitenev, N. B.; Stroscio, J. A. Microscopic Polarization in Bilayer Graphene. Nature Physics 2011, 7 (8), 649–655.

\bibitem{sts4} Li, G.; Andrei, E. Y. Observation of Landau Levels of Dirac Fermions in Graphite. Nature Physics 2007, 3 (9), 623–627.

\bibitem{op-sp2} Orlita, M.; Faugeras, C.; Borysiuk, J.; Baranowski, J. M.; Strupinski, W.; Sprinkle, M.; Berger, C.; de Heer, W. A.; Basko, D. M.; Martinez, G.; et al. Magneto-Optics of Bilayer Inclusions in Multilayered Epitaxial Graphene on the Carbon Face of SiC. Phys. Rev. B 2011, 83 (12), 125302.

\bibitem{op-sp3} Plochocka, P.; Faugeras, C.; Orlita, M.; Sadowski, M. L.; Martinez, G.; Potemski, M.; Goerbig, M. O.; Fuchs, J.-N.; Berger, C.; de Heer, W. A. High-Energy Limit of Massless Dirac Fermions in Multilayer Graphene Using Magneto-Optical Transmission Spectroscopy. Phys. Rev. Lett. 2008, 100 (8), 087401.

\bibitem{op-sp4} Plochocka, P.; Solane, P. Y.; Nicholas, R. J.; Schneider, J. M.; Piot, B. A.; Maude, D. K.; Portugall, O.; Rikken, G. L. J. A. Origin of Electron-Hole Asymmetry in Graphite and Graphene. Phys. Rev. B 2012, 85 (24), 245410.

\bibitem{op-sp5} Nicholas, R. J.; Solane, P. Y.; Portugall, O. Ultrahigh Magnetic Field Study of Layer Split Bands in Graphite. Phys. Rev. Lett. 2013, 111 (9), 096802.

\bibitem{tbm} Chen, S.-C.; Wu, J.-Y.;Lin, C.-Y.;Lin, M.-F. Theory of Magnetoelectric Properties of 2D Systems. IOP Publishing 2017, 2053-2563, ISBN: 978-0-7503-1674-3.

\bibitem{approx} Lin, M. F.; Shung, K. W.-K. Plasmons and Optical Properties of Carbon Nanotubes. Phys. Rev. B 1994, 50 (23), 17744–17747.

\end{references}
\end{document}